# Spectroscopic signature of the Stark-shifted Tamm-type surface state of La(0001)

Dominik Schreyer, Howon Kim[*], and Roland Wiesendanger[*]

Department of Physics, University of Hamburg, D-20355 Hamburg, Germany

Corresponding authors(*):

hkim@physnet.uni-hamburg.de

wiesendanger@physnet.uni-hamburg.de

**Abstract:**

**We have studied the Tamm-type surface state of La(0001) by tunneling spectroscopy within a wide range of tunneling currents from 0.1 nA to 8000 nA, thereby tuning the electric-field strength in a tip-vacuum-sample tunnel junction. We observe a significant shift of the unoccupied Tamm-type surface state toward the Fermi energy with increasing electric-field strength, accompanied by a broadening of the width of the resonance peak indicating a decrease of the surface-state lifetime. Our experimental results are contrary to previous reports for Stark-shifted Shockley-type surface states of noble metal (111) surfaces.**

## 1. Introduction

In scanning tunneling microscopy/spectroscopy (STM/S), a significant electric field is present between the STM tip and the sample surface, which can cause a Stark shift of the electronic states observed in the tunneling spectra. The Stark effect is well known in semiconductor physics, where it is utilized to tune electronic states, for example shifting excitons in layered quantum wells (quantum confined Stark effect) [1] and quantum dots [2], as well as optical excitation lines of single defects in diamond [3]. When performing STM/S experiments on metal surfaces with relatively low tunneling currents (0.1 – 10 nA), the Stark effect is often regarded as negligible concerning its influence on the spectroscopic measurements. However, Limot *et al.* [4] and Kröger *et al.* [5] have reported a Stark shift of occupied Shockley-type surface states of Ag(111), Au(111) and Cu(111) as a general feature in STS measurements. Shockley surface states [6] are derived from *s*- and *p*-electronic bands in the gap of the bulk band structure. In contrast to the noble metal (111) surfaces, rare earth metal (0001) surfaces are known to exhibit Tamm-type surface states [7] originating from *d*-band electrons. They have been investigated previously by spin-averaged and spin-resolved STS [8-11] using conventional tunneling parameters. Here, we report on a tunneling spectroscopy study of the unoccupied Tamm-type surface state of La(0001) performed within a wide range of tunneling currents. Since the electric field direction changes when performing STM/S measurements with positive sample bias, thereby probing electronic states with energies $E > E_F$, our investigation provides complementary information about the surface state's Stark shifting behavior compared to previous STM/S studies of occupied surface states. Recently, there has been an increasing interest in lanthanum as elemental superconductor [12,13], and due to the exceptional superconducting properties of La-hydrides [14,15], which motivated the current research of its electronic states. To the best of our knowledge, there have been no previous studies of the Stark effect for unoccupied electronic states of metal surfaces. A fundamental question, which arises in this context is whether unoccupied electronic states will be affected similarly as occupied electronic states by the presence of a strong electric field being reversed regarding its direction. Moreover, the lifetime of surface states under the influence of a local electric field is an interesting topic, which has not been addressed so far due to the absence of experimental signatures. Theoretically, a linear increase of the decay rate as a function of Stark-shifted energy due to electron-electron or electron-phonon interactions at the bottom of the surface-state band as well as for image potential states just above the surface was predicted based on many-body calculations [16, 17]. For the (111) surfaces of noble metals, however, the influence of the local electric field on the lifetime of the surface states was found to be negligible in STM experiments [4, 5], in stark contrast to the case of the image potential states [4, 16, 17].

In the present study, we monitored the influences of the local electric field on the Tamm-type surface state of La(0001) in local tunneling spectra by varying the set-point tunneling current from 0.1 pA to 8.0 μA in an STM junction. By reducing the tip-surface distance, i.e. increasing the local electric field, we reveal a significant energy shift of the Tamm-type surface state toward the Fermi energy as well as a sizable increase of the surface-state lifetime based on the analysis of the line-shape in the tunneling spectra. We attribute these observations to a significant Stark effect on the unoccupied Tamm-type surface state of La(0001) under the influence of a local electric field, including electric field-driven modifications of the electron-electron or electron-phonon interactions at this surface. Our study therefore provides strong experimental evidence for a remarkable influence of the electric field on the efficiency of inelastic scattering channels on the La(0001) surface.

## 2. Experimental details

The experiments were performed using a low-temperature STM system (USM 1300, Unisoku, Japan) in an ultra-high vacuum (UHV) environment at a temperature of 1.7 K. A La(0001) film of >50 nm thickness was obtained by evaporating Lanthanum onto a clean Re(0001) substrate with subsequent annealing at 700 °C for 15 minutes. Tunneling spectroscopy measurements were performed using a PtIr tip coated with La to avoid material ambiguity. A lock-in technique with open feedback loop has been applied for the spectroscopic measurements using a modulation voltage of 3 mV and a frequency of 1.175 kHz. After fixing the spatial position of the STM tip on a defect-free region of the La(0001) surface, we measured the differential tunneling conductance $dI/dV$ by sweeping the sample bias at given stabilization tunneling conditions (tunneling current, $I_T$ and sample bias, $V_S$). Different tunneling currents lead to a different distance between tip and sample, thereby changing the electric field strength in the vacuum tunnel junction. Since the *dI/dV* spectra directly reflect the local density of states (LDOS) of the surface, the influence of the local electric field on the surface electronic states was studied by collecting tunneling spectra *dI/dV(V)* as a function of the tunneling current $I_T$.

## 3. Results and discussion

Fig. 1 illustrates the principle of our distance-dependent vacuum tunneling experiments. The surface state of La(0001) is energetically located above the Fermi energy ($E_F$). Therefore, the applied sample bias voltage has to be of positive sign and the electrons have to tunnel from the STM tip into the sample (Fig. 1(a)), while the local electric field is pointing in the direction of the STM tip (Fig. 1(b)). By increasing the set-point (or stabilization) tunneling current, the tip-sample distance decreases, leading to an increasing electric field strength (Fig. 1c).

Fig. 2(a) shows measured tunneling spectra on La(0001) for four different stabilization values of the tunneling current ($I_T$) ranging from 0.1 nA to 8000 nA, while the sample bias voltage was ramped between -300 meV and +300 meV. The blue curve in Fig. 2(a) shows the tunneling spectrum *dI/dV(V)* for $I_T$ = 0.1 nA. The Tamm-type surface state of La(0001) can be observed at $E$ = 106 meV, in agreement with previously published tunneling spectroscopy data [11-13]. The green, orange and red curves show the tunneling spectra obtained for $I_T$ = 100 nA, 1000 nA, and 8000 nA, respectively. The Tamm-type surface state of La(0001) experiences a total shift of 45.9 meV towards lower energy for the tunneling current range chosen for this experiment, resulting in a shift of 43.3 % with respect to the peak position for $I_T$ = 0.1 nA. Near the Fermi energy the superconducting gap of the La sample is visible in all tunneling spectra. Changing the tunneling current and therefore the tip-surface distance as well as the electric field in the tunnel junction does not induce a detectable energy-shift of the superconducting coherence peaks. The *dI/dV* spectra were normalized to the conductance at $E$ = -200 meV, where the *dI/dV(V)* spectra appear to be nearly constant in order to directly compare the spectral features resulting from different tunneling current set-points.

Fig. 2(b) presents the spectral evolution of the Tamm-type surface state of La(0001) in terms of the peak position of the resonance peak ($E_0$, black) and its full width at half maximum (FWHM, red) in dependence of the tunneling current. For deriving quantitative values for $E_0$ and the FWHM of asymmetric resonance peaks as shown in Fig. 2(a), all tunneling spectra obtained for $I_T$ = 0.1 nA – 8000 nA have been fitted using a bi-gaussian function:

$$y = y_0 + H\, e^{-0.5\left(\frac{x-x_c}{w_1}\right)^2} \quad (x < x_c) \qquad (1)$$

$$y = y_0 + H\, e^{-0.5\left(\frac{x-x_c}{w_2}\right)^2} \quad (x \geq x_c) \qquad (2)$$

with $y_0$ being the baseline, $H$ the amplitude, $x_c$ the center of the peak, $w_1$ ($w_2$) the half-width at half maximum (HWHM) for the left (right) half of the peak. We obtained the FWHM by summing $w_1$ and $w_2$. Interestingly, in addition to the energy-shift of the surface state peak, we observe a sizable broadening of the width as the tunneling current is increased, i.e. by enhancing the electric field strength.

In general, the width of a surface-state peak originating from the LDOS of the quasiparticles in the surface-state band can be approximately attributed to the lifetime of the quasiparticles based on the relationship: $\Gamma \propto 1/\tau$, where $\Gamma$ is the FWHM and $\tau$ the lifetime of the quasiparticles [11]. A rise in FWHM therefore corresponds to a decrease of the surface state's quasiparticle lifetime. Theoretically, the lifetime of the quasiparticle excitation in metallic surface bands can be attributed to the decay rate of an excitation, which can be obtained from the expectation value for the imaginary part of the electron self-energy [16]. By considering the local electric field in the STM junction, the decay rate of the hole-like excitation at the bottom of the occupied Shockley surface-state band on Ag(111) has shown a linear increase as the surface state is shifted to larger binding energies. To elucidate the dynamics related with the Tamm-type surface state of La(0001), we mapped the values of ($E_0$, $\Gamma$) for the different tunneling currents as shown in Fig. 2(c). We find a clear linear increase of $\Gamma$ as $E_0$ is Stark-shifted to lower energy. Remarkably, the rate of increase is considerably larger with $|d\Gamma/dE_0| = 2.14$ compared to $|d\Gamma/dE_0| = 0.05$ for the Shockley-type surface state of Ag(111) [16] and $|d\Gamma/dE_0| = 0.037$ for the Cu(001) $n = 1$ image state [17]. Such a large decay rate of a Tamm-type surface state has not been reported previously.

It is, however, necessary to consider the possible influence of the proximity of the STM tip, being very close to the surface at high tunneling currents, on the surface state lifetime. When an STM junction is operated away from the tunneling regime, the spectral line-shape of an electronic state can be modified due to the considerable interaction between tip and surface regardless of the presence of a local electric field [18]. To evaluate this possibility, we have plotted the dependence of the tunneling current on a logarithmic scale as a function of tip displacement for our tunneling spectroscopy measurement on La(0001) using a La-coated Pt/Ir-tip. As can be seen clearly from this data, the tunneling current exhibits the characteristic exponential dependence on tip-surface distance within the whole range of currents from 0.1 nA up to 8000 nA, corresponding to a total tip displacement of 7.3 Å. This observation indicates that the STM junction remains in the tunneling regime during our experiments, far away from the contact regime between the two electrodes. Therefore, we can safely rule out an influence of the direct interaction between tip and surface onto the spectral evolution of the measured tunneling spectra.

In the work of Limot et al. [4], the effect of different tip work functions $\Phi_{tip}$ on the energetic position of the surface state peak vs. tip displacement has been considered and found to be significant. For our measurements presented in Fig. 2, a La-coated Pt/Ir tip ($\Phi_{tip}$ = 3.5 eV [19]) has been used to keep the difference in work function between STM tip and sample as low as possible. For comparison, we have performed additional tunneling spectroscopy measurements with an uncoated Pt/Ir tip ($\Phi_{tip}$ = 5.5 eV [20]) and a Cr tip ($\Phi_{tip}$ = 4.5 eV [19]). As shown in Fig. 3(a) and 3(b) the Stark effect on the surface state peak results in a very similar shift towards lower energy and a broadening of the peak with increasing tunneling current when measuring with these two other types of STM tips. Fig. 3(c) shows the surface state peak position *vs.* tip displacement for all three series of tunneling spectroscopy measurements performed with the three different STM tips mentioned above. Although the tunneling current range is smaller (1 nA – 1000 nA) for the measurement series with the Cr- and Pt/Ir-tip in comparison to the measurement series with the La-coated tip (0.1 nA – 8000 nA), the different data sets show the same trend despite the strong difference of up to 2.0 eV in the tip work functions (uncoated vs. La-coated Pt/Ir tip).

The observed shift in energy of the Tamm-type surface state of La(0001) can be explained by the Stark effect due to the electric field present between tip and sample, similar to the behavior of Shockley-type surface states of noble metal (111) surfaces [4, 5]. The major difference lies in the fact that La(0001) exhibits an unoccupied instead of an occupied surface state. This means that the direction of the tunneling current and the electric field between tip and sample is inverted. However, the results presented in Fig. 2 show clearly, that the direction of energy shift for the Tamm-type surface state of La(0001) towards lower energies is rather the same compared to the Shockley-type surface state [4]. On the other hand, there are several differences between our present study and the earlier reports for noble metal (111) surfaces.

In the work of Limot *et al.* [4] it is reported that the tunneling current shows an exponential distance dependence only until a certain threshold of about 2000 nA. As a reason for that behavior a reversible deformation of the STM tip at very small tip-sample distances has been proposed. In contrast, an exponential distance dependence of the tunneling current is observed even for the highest set-point currents up to 8000 nA in the present experiment, thus no indications for an elastic tip deformation or a jump-to-contact between tip and surface have been found (Fig. 2c).

As can be seen from Fig. 3, different tip work functions (with a difference of up to 2.0 eV) lead to a similar distance-dependent behavior of the Tamm-type surface state of La(0001) for all three STM tips used. These experimental results are in contrast to the results of the theoretical model calculations for the Ag (111) surface state behavior [4]. In their report, the binding energy shift of the Shockley-type surface state showed a significant dependency on the work function, resulting even in a positive Stark shift.

While no increase in the FWHM of the surface state peak's width was observed for Ag(111) [4], our experimental results on La(0001) clearly show an increase of the FWHM with decreasing tip-sample distance (Fig. 2b). Based on the relation between the FWHM and the surface state's lifetime, a strong decrease of the lifetime upon decreasing the tip-sample distance is evident from our data. The large decay rate can be explained by the increased scattering of quasiparticles due to electron-electron and electron-phonon interactions under the influence of the local electric field. According to an analysis of the quasiparticle interference pattern for the Tamm-type surface state of La(0001) together with the surface band structure based on DFT calculations [11, 21], there is an overlap of the surface band with the bulk band along the ΓK direction. Although the dominant scattering channel is the intraband scattering in the La(0001) surface-state, we expect that the opening of the interband scattering channels between surface and bulk band under the influence of a local electric field may be responsible for the large decay rate of the quasiparticle excitations on La(0001).

To explain the observed Stark shifting behavior of surface states as probed by tunneling spectroscopy, the STM-tip induced dipole potential at the sample surface might play an important role. Indeed, the influence of a surface dipole moment induced by the STM tip on the Stark shift of Au/Fe(100) quantum well (QW) states has been reported [22]. By including the induced dipole moment in a 1D potential model [4, 23], an additional phase shift of the QW state wave function appears. As a result, larger energy shifts for the occupied and unoccupied states were obtained compared to results based on simpler types of 1D model calculations. The tip-induced dipole potential was not included in the potential model of Limot *et al.* [4] and might explain the discrepancy between their theoretical and experimental results. The studies on QW states [22] revealed a negative Stark shift for unoccupied electronic states, in agreement with our results for La(0001) presented here, while the energy shift for occupied states was found to be positive, in contrast to the previously reported results for Ag(111) [4].

## 4. Conclusion

In summary, we have studied the Tamm-type surface state of La(0001) by tunneling spectroscopy as a function of tunneling current and therefore tip-sample distance. The effect of the increasing electric field strength with decreasing tip-sample separation could be observed by a Stark shift of the surface state's energy position in the tunneling spectra. It was found that the surface state peak of La(0001) experiences an energetic downshift, similar to the occupied Shockley-type states of noble metal (111) surfaces, with increasing electric field strength. Additionally, based on the spectroscopic line-shape analysis, we have addressed the dynamics of a Tamm-type surface state for the first time, reflecting a remarkably large decay rate of the quasiparticle excitations under the influence of a local electric field. One might expect that for occupied and unoccupied surface states, which are probed with an opposite bias voltage polarity in tunneling spectroscopy and therefore a different direction of the local electric field, the Stark shift of the surface state peaks corresponding to occupied and unoccupied states might change its direction in energy as well. However, this is obviously not the case. Moreover, we find a negligible dependence of the chosen STM tip material on the measured data of the surface state peak position vs. tip-sample displacement in our experiment, meaning that the effect of tip work function and therefore contact potential differences is negligible as well. Since we found a pure exponential behavior of the tunneling current as a function of tip displacement, we believe that our results are characteristic for the tunneling regime and not influenced by other spurious effects originating from elastic tip deformations or jump-to-contact behavior. However, further theoretical studies based on first-principles electronic structure calculations are needed in order to understand the Stark shifting behavior of the unoccupied La(0001) surface state quantitatively.

**Acknowledgments**

We thank L. Rózsa, E. Simon and J. Wiebe for fruitful discussions. This work was supported by the European Research Council via project no. 786020 (ERC Advanced Grant ADMIRE).

**References**

[1] D. A. B. Miller, D. S. Chemla, T. C. Damen, A. C. Gossard, W. Wiegmann, T. H. Wood, C. A. Burrus, Phys. Rev. Lett. **53**, 2173 (1984).

[2] V. I. Klimov, S. A. Ivanov, J. Nanda, M. Achermann, I. Bezel, J. A. McGuire, A. Piryatinski, Nature **447**, 441 (2007).

[3] P. Tamarat, T. Gaebel, J. R. Rabeau, M. Khan, A. D. Greentree, H. Wilson, L. C. L. Hollenberg, S. Prawer, P. Hemmer, F. Jelezko, J. Wrachtrup, Phys. Rev. Lett. **97**, 083002 (2006).

[4] L. Limot, T. Maroutian, P. Johansson, and R. Berndt, Phys. Rev. Lett. **91**, 196801 (2003).

[5] J. Kröger, L. Limot, H. Jensen, R. Berndt, and P. Johansson, Phys. Rev. B **70**, 033401 (2004).

[6] W. Shockley, Phys. Rev. **56**, 317 (1939).

[7] I. Tamm, Phys. Z. Sowjetunion **1**, 733 (1932).

[8] M. Bode, M. Getzlaff, and R. Wiesendanger, Phys. Rev. Lett. **81**, 4256 (1998).

[9] M. Bode, M. Getzlaff, A. Kubetzka, R. Pascal, O. Pietzsch, and R. Wiesendanger, Phys. Rev. Lett. **83**, 3017 (1999).

[10] L. Berbil-Bautista, S. Krause, M. Bode, and R. Wiesendanger, Phys. Rev. B **76**, 064411 (2007).

[11] D. Wegner, A. Bauer, Y. M. Koroteev, G. Bihlmayer, E. V. Chulkov, P. M. Echenique, G. Kaindl, Phys. Rev. B **73**, 115403 (2006).

[12] P. Löptien, L. Zhou, A. A. Khajetoorians, J. Wiebe, and R. Wiesendanger, J. Phys.: Condens. Matter **26**, 425703 (2014).

[13] P. Löptien, L. Zhou, A. A. Khajetoorians, J. Wiebe, and R. Wiesendanger, Surf. Sci. **643**, 6 (2016).

[14] A. P. Drozdov, P. P. Kong, V. S. Minkov, S. P. Besedin, M. A. Kuzovnikov, S. Mozaffari, L. Balicas, F. F. Balakirev, D. E. Graf, V. B. Prakapenka, E. Greenberg, D. A. Knyazev, M. Tkacz & M. I. Eremets, Nature **569**, 528 (2019).

[15] M. Somayazulu, M. Ahart, A. K. Mishra, Z. M. Geballe, M. Baldini, Y. Meng, V. V. Struzhkin, and R. J. Hemley, Phys. Rev. Lett. **122**, 027001 (2019).

[16] M. Becker, S. Crampin, R. Berndt, Phys. Rev. B **73**, 081402(R) (2006).

[17] S. Crampin, Phys. Rev. Lett. **95**, 046801 (2005).

[18] N. Néel, J. Kröger, L. Limot, K. Palotas, W. A. Hofer, R. Berndt, Phys Rev. Lett. **98**, 016801 (2007).

[19] W. M. Haynes, D. R. Lide, T. J. Bruno, *CRC Handbook of Chemistry and Physics* (CRC Press, 2015).

[20] E. T. Yu, K. Barmak, P. Ronsheim, M. B. Johnson, P. McFarland, J.-M. Halbout, J. Appl. Phys. **79**, 2115 (1996).

[21] H. Kim, L. Rózsa, D. Schreyer, E. Simon, and R. Wiesendanger (unpublished).

[22] S. Ogawa, S. Heike, H. Takahashi, T. Hashizume, Phys. Rev. B **75**, 115319 (2007).

[23] E. V. Chulkov, V. M. Silkin, P. M. Echenique, Surf. Sci. **437,** 330 (1999).

Figure 1

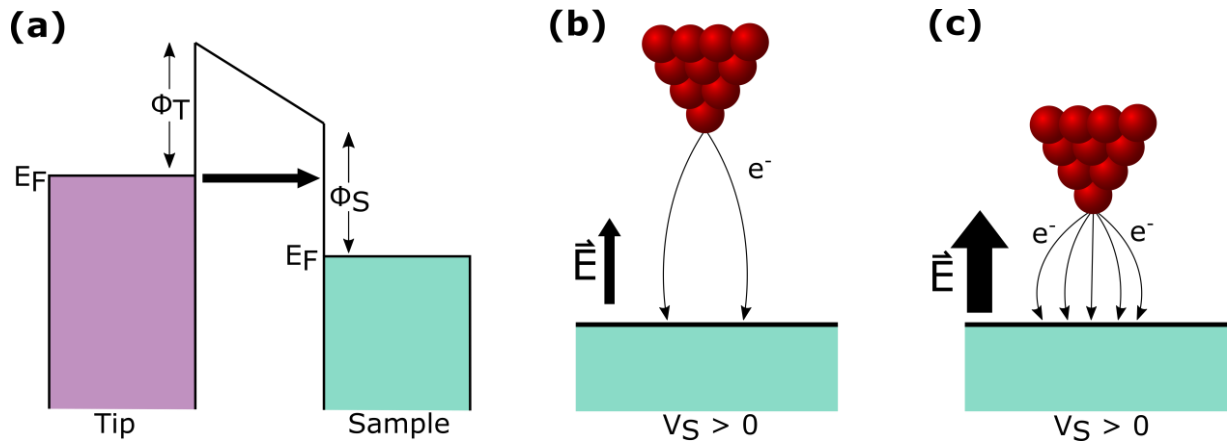

**Figure 1. Schematics of distance-dependent tunneling spectroscopic experiments on La(0001). (a)** Energy diagram of the tunnel junction. The unoccupied surface state of La(0001) is probed with a positive sample bias voltage, resulting in electron tunneling from the STM tip to the sample (thick arrow). $E_F$ denotes the Fermi energy, while $\Phi_T$ and $\Phi_S$ are the work functions of tip and sample. **(b)** Spatial diagram of the tunnel junction. The local electric field is pointing from the sample towards the STM tip. **(c)** At a higher set-point tunneling current the tip-surface distance decreases while the electric field strength increases.

Figure 2

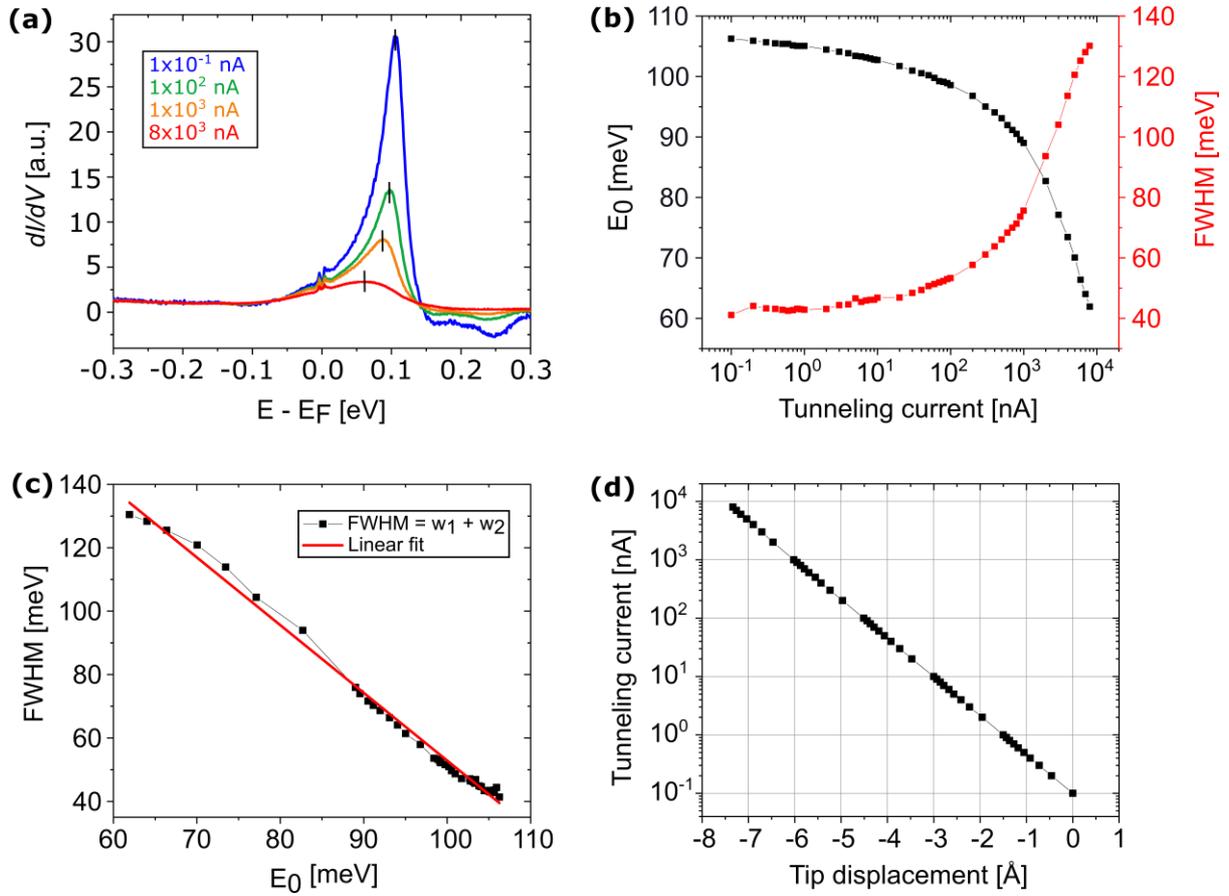

**Figure 2. Spectral evolution of the Tamm-type surface state of La(0001) depending on the stabilization current.** (**a**) *dI/dV* spectra showing the Tamm-type surface state of La(0001) as resonance peaks and its evolution with increasing stabilization current ($I_T$ = 0.1 nA – 8000 nA, $V_S$ = -300 meV). The black marks represent the energy position of each surface state peak maximum. The signature of the superconducting gap of La can be seen around zero energy. (**b**) Surface state peak positions ($E_0$, black) and FWHM of each peak (red) plotted as a function of the tunneling current. (**c**) FWHM plotted against the surface state peak positions ($E_0$). A linear fit of the data points with a slope of 2.14 is presented in red. (**d**) Semi-logarithmic plot of the tunneling current vs. tip displacement. A linear dependence of the tunneling current on a logarithmic scale, implying a characteristic exponential behavior of the tunneling current with tip displacement, is observed for the whole range of currents (0.1 nA – 8000 nA). Zero displacement is corresponding to the initial position of the tip relative to the surface for $I_T$ = 0.1 nA. Negative displacements at higher tunneling currents correspond to a tip movement toward the surface.

**Figure 3**

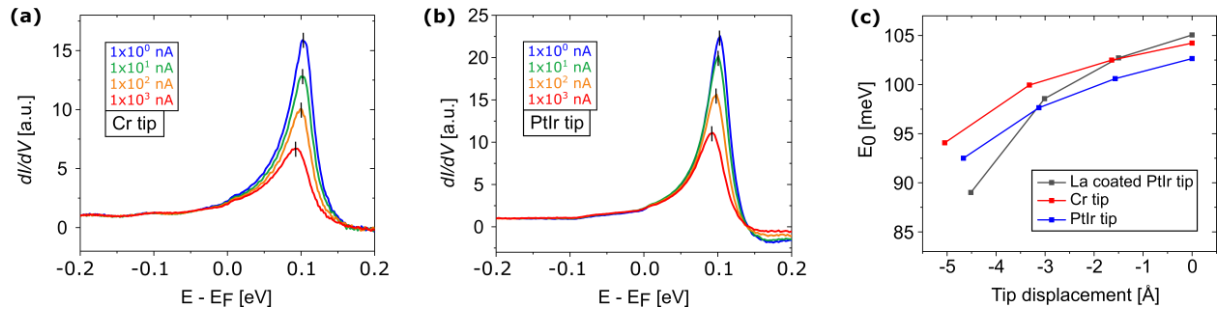

**Figure 3. Surface state behavior of La(0001) as a function of tunneling current as measured with a Cr- and PtIr tip**. (**a**) *dI/dV* spectra of the La(0001) surface as measured with a Cr tip ($I_T$ = 1 nA – 1000 nA, $V_S$ = 400 meV), normalized at -200 meV. (**b**) Similar set of *dI/dV* spectra as measured with a Pt/Ir tip ($I_T$ = 1 nA – 1000 nA, $V_S$ = 600 meV), normalized at -200 meV. (**c**) Surface state peak position ($E_0$) vs. tip displacement derived from the tunneling spectra measured with three different tips. The four data points corresponding to measurements with Cr- and PtIr tips were extracted from the four *dI/dV* spectra presented in (**a**) and (**b**), respectively. For comparison, the four points representing data obtained with a La-coated tip are derived from Fig. 2.